\documentclass[12pt]{article}
\usepackage{amssymb}
\usepackage{amsbsy}
\usepackage{epsfig}
\usepackage{verbatim}
\makeatletter \@addtoreset{figure}{section}
\def\thefigure{\thesection.\@arabic\c@figure}
\def\fps@figure{h, t}
\@addtoreset{table}{bsection}
\def\thetable{\thesection.\@arabic\c@table}
\def\fps@table{h, t}
\@addtoreset{equation}{section}

\newtheorem{corollary}{Corollary}[section]
\newtheorem{definition}{Definition}[section]
\newtheorem{theorem}{Theorem}[section]
\newtheorem{proposition}{Proposition}[section]

\newtheorem{lemma}{Lemma}[section]
\newtheorem{remark}{Remark}[section]
\newtheorem{remarks}[remark]{Remarks}

\def\bd{\begin{definition}}
\def\ed{\end{definition}}
\def\bt{\begin{theorem}}
\def\et{\end{theorem}}
\def\bp{\begin{proposition}\rm}
\def\ep{\end{proposition}}
\def\bc{\begin{corollary}}
\def\ec{\end{corollary}}
\def\bl{\begin{lemma}\em}
\def\el{\end{lemma}}
\def\be{\begin{equation}}
\def\ee{\end{equation}}
\def\br{\begin{remark}\rm\small}
\def\er{\end{remark}}
\def\brs{\begin{remarks}.\\ \rm\
\begin{enumerate}}
\def\ers{\end{enumerate}\end{remarks}}
\def\bea{\begin{eqnarray}}
\def\eea{\end{eqnarray}}

\def\Tr{\mathrm {Tr}}
\def\tr{\mathrm {tr}}

\def\diag{\mathrm {diag}}

\def\&{&{\hskip -20pt}}

\def\pb{\mathbf{p}}

\def\nchi{\hbox{\raise 2.5pt\hbox{$\chi$}}}
\date{}
\begin{document}
\baselineskip 16pt
\medskip
\begin{center}
\begin{Large}\fontfamily{cmss}
\fontsize{17pt}{27pt}
\selectfont
\textbf{Products of random Hermitian matrices and brickwork Hurwitz numbers}
\end{Large}\\
\bigskip
\begin{large}
 {Chuanzhong Li$^{1}$ and A. Yu. Orlov}$^{1,2}$
 \end{large}
\\
\bigskip
\begin{small}
$^{1}${\em College of Mathematics and Systems Science, Shandong University of Science and Technology,\\
Qingdao, 266590, Shandong, P. R. China\\
 e-mail: lichuanzhong@sdust.edu.cn }\\
$^{2}${\em Nonlinear Wave Processes Laboratory, \\
Oceanology Institute, 36 Nakhimovskii Prospect,
Moscow 117851, Russia\\
 e-mail: orlovs@ocean.ru } \\
\end{small}
\end{center}
\bigskip


\begin{abstract}
We consider products of $n$ random Hermitian matrices which generalize the
one-matrix model and show its relation to Hurwitz numbers
which count ramified coverings of certain type. Namely, these
Hurwitz numbers  count $2k$-fold ramified coverings of the Riemann sphere with arbitrary ramification type over $0$ and $\infty$ and ramifications related to the partition $(2^k)$ (``brickworks'' - involutions without fixed points) elsewhere.
\end{abstract}
\smallskip

\tableofcontents

\section{Introduction}

The famous one-matrix model is a generalization of the matrix models of quantum gravity
\cite{BK},\cite{GM} and is the most studied and popular models of random matrices.
Its partition function is
\be\label{old}
Z_N(p)=\int_{\Sigma_N} e^{N\sum_{m=1}^\infty \frac{p_m}{m}\Tr\left(H^m\right)}d\mu(H)
\ee
where $\Sigma_N$ is the space of $N\times N$ Hermitian matrices and $d\mu(H)$ is the Gaussian measure on $\Sigma$:
$$
d\mu(H)=C_Ne^{-\frac N2 \tr HH^\dag}dH,\quad dH=\prod_{i=1}^N dH_{ii}\prod_{i<j}^N d^2 H_{i,j},\quad
\int_{\Sigma_N} d\mu(H)=1
$$
The space $\Sigma$ and the measure $d\mu$ is called unitary ensemble, see \cite{Mehta}.

Here we consider the multi-matrix models
\be\label{new}
Z^{(n)}_N(p)=\int_{\Sigma_N^{\times n}}
e^{N\sum_{m=1}^\infty \frac{p_m}{m}\Tr\left((H_1\cdots H_n)^m\right)}\prod_{i=1}^nd\mu(H_i)
\ee
and
\be\label{new'}
Z^{(n)}_N(p,C)=\int_{\Sigma_N^{\times n}}
e^{N\sum_{m=1}^\infty \frac{p_m}{m}\Tr\left((H_1C_1\cdots H_nC_n)^m\right)}\prod_{i=1}^nd\mu(H_i)
\ee
where we call arbitrary matrices $C=\{C_1,\dots,C_n\}$ source matrices. In case each $C_i$ is identity
matrix we get (\ref{new}). Each of $C_i$ can be chosen to be random to construct new model.

\section{Matrix models (\ref{new}), (\ref{new'})}

\subsection{Preliminaries}

To evaluate the integral (\ref{new}) we need the following
\bp
 Let $A$ and $B$ be $N\times N$ complex matrices. Then
\be\label{Prop1}
C_N\int_{\mathbb{U}_N} s_\lambda(UAU^\dag B) d_*U=\frac{s_\lambda(A)s_\lambda(B)}{s_\lambda(I_N)}
\ee
where $d_*U$ is the Haar measure on $\mathbb{U}_N$ and $C_N$ is chosen from the normalization
$C_N\int_{\mathbb{U}_N} d_*U=1$ (and in what follows we include it to the definition of $d_*U$) and where $s_\lambda$ is the Schur function labeled by
a partition $\lambda$, see \cite{Mac}.
\ep

It can be known, however, since we did not see the written proof of (\ref{Prop1}) we need to present it below.
The proof consists of few simple steps (i)-(iii):

\,

(i) Firstly, we note that for $A,B\in\Sigma $, where $\Sigma$ is the space of Hermitian matrices
the identity (\ref{Prop1}) is known, see
 Example 3  in Section 5 of Chapter VII in the textbook \cite{Mac}, the page 445 (and also page 441 where $\Sigma$ is defined).

 (ii) Then, due to the invariant property of the Haar measure we can replace $A$ and
 $B$ by diagonal matrices of eigenvalues $X=\diag(x_1,\dots,x_N)$ and $Y=\diag(y_1,\dots,y_N)$ respectively, where $x_a,y_a$ are real.  Then, since the both sides of (\ref{Prop1}) are polynomial in each variable
 $x_a,y_a$ ($a=1,\dots,N$) we can analiticaly continue both sides for complex $x_a,y_a$. Thus,
 we get
 $$
\int_{\mathbb{U}_N} s_\lambda(UXU^\dag Y) d_*U=\frac{s_\lambda(X)s_\lambda(Y)}{s_\lambda(I_N)}
 $$
 where $X$ and $Y$ are complex diagonal\footnote{Using the Cauchy-Littlewood relation the this was used in \cite{O-new} to re-derive
 the determinantal answer for the HCIZ integral \cite{HarishChandra},\cite{ItzyksonZuber}
 $\int e^{\tr UAU^\dag B}d_*U$
 for normal matrices $A$ and $B$.}.
 
 Consider strictly upper-triangular matrices
\be\label{Delta}
\Delta^{(A)}_{a,b}=\Delta^{(B)}_{a,b}=0,\quad a\ge b.
\ee

Suppose that $\tilde A$ and $\tilde B$ have the form
\be\label{Schur-decomposition}
\tilde A=U_1(X+\Delta^{(A)})U_1^\dag,\quad
\tilde B=U_2(Y+\Delta^{(B)})U_2^\dag,\qquad
U_i\in \mathbb{U}_N,\quad i=1,2
\ee
where $X=\diag\{x_1,\dots,x_N\}$ and $Y=\diag\{y_1,\dots,y_N\}$ are complex diagonal matrices, and $\Delta^{(i)}$ are strictly upper-triangular, see (\ref{Delta}).
Now $\tilde A$ and $\tilde B$ are arbitrary complex matrices and (\ref{Schur-decomposition})
are their Schur decomposition.
We have $s_\lambda(\tilde A)=s_\lambda(X)$ and $s_\lambda(\tilde B)=s_\lambda(Y)$.
Let us prove that
\be
\int_{\mathbb{U}_N} s_\lambda(UAU^\dag B) d_*U=
\int_{\mathbb{U}_N} s_\lambda(UXU^\dag Y) d_*U.
\ee

To verify it let us use a result of B.Collins (see \cite{Collins} and \cite{Collins2}).
Let
\be
{\cal U}_N({\bf a},{\bf a}',{\bf b},{\bf b}'):=
 U_{a_1,b_1}\cdots U_{a_d,b_d}(U^\dag)_{b'_1,a'_1}\cdots (U^\dag)_{b'_{d'},a'_{d'}}
\ee
which is a monomial in the entries of matrices $U\in\mathbb{U}_N$ and it's inverse one
labeled, respectively, by sets ${\bf a},{\bf b}$ and ${\bf a}',{\bf b}'$.

Then, the nice result of \cite{Collins} is
\be\label{collins}
\int_{\mathbb{U}_N}{\cal U}_N({\bf a},{\bf a}',{\bf b},{\bf b}')d_*U
=\delta_{d,d'}\sum_{\sigma,\tau\in S_d}{Wg}_N(\tau\sigma^{-1})\prod_{k=1}^c\delta_{a_k,a'_{\sigma(k)}}\delta_{b_k,b'_{\tau(k)}}
\ee
where $S_c$ is the symmetric group and $Wg_N$ is some function on $S_c$. Let us note that the integral in the left hand side
of (\ref{collins}) was also evaluated in other works, see \cite{Morozov-U-int},\cite{Shatashvili}, however it is (\ref{collins}) we need in.

\br
$Wg_N$ is called Weingardten's function which depends only on the cycle structure  of the permutation $\sigma$ (say, given by partition $\mu=\mu(\sigma)$) and on $N$.
We do not need it's explicit form, however let us write down to show it's relation to
tau-functions 
\be
Wg_N(\sigma) =Wg_N\left(\mu(\sigma)\right) =\sum_{\lambda\atop |\lambda|=|\mu(\sigma)|=d}\frac{s_\lambda(\pb_\infty)\chi_\lambda(\mu(\sigma))}{(N)_\lambda}
\ee
\be
=\int_{\mathbb{U}_N} U_{1,1}\cdots U_{d,d}(U^\dag)_{\sigma(1),1}\cdots (U^\dag)_{\sigma(d),d}
d_*U
\ee
where $\chi_\lambda(\sigma)=\chi_\lambda(\mu(\sigma))$ is the character of the representation $\lambda$ of the symmetric group $S_{|\lambda|}$ evaluted on the cycle class $\mu(\sigma)$, where
$\pb_\infty=(1,0,0,\dots)$  and where
\be\label{cont-prod}
(N)_\lambda :=\prod_{(i,j)\in\lambda} (N+j-i)
\ee
is called the content product.
Then it follows that 
\be
\sum_{\mu}  Wg_N(\mu(\sigma))\frac{\pb_\mu}{z_\mu} = \sum_\lambda 
\frac{s_\lambda(\pb_\infty)s_\lambda(\pb)}{(N)_\lambda}
=\tau_N(\pb)
\ee
is the example of the modified KP tau function and $N$ and $\pb=(p_1,p_2,\dots)$ play the role of higher times. Here $z_\mu=\prod_{i}i!i^{m_i}$, where each part $i$ ($i=1,2,\dots$ enters 
$m_i$ times to the partition $\mu$).
\er

In view of (\ref{collins}) in case $d=d'$ it is natural to introduce two balance numbers
(or, balance functions defined on monomials terms where $d=d'$)
\be
{\cal A}\left({\cal U}_N({\bf a},{\bf a}',{\bf b},{\bf b}')\right)=\sum_{i=1}^d a_i-\sum_{i=1}^{d'} a'_i
\ee
and
\be
{\cal B}\left({\cal U}_N({\bf a},{\bf a}',{\bf b},{\bf b}'\right)=\sum_{i=1}^{d} b_i-\sum_{i=1}^{d'} b'_i.
\ee

In what following we omit the dependence of $N,{\bf a},{\bf a}',{\bf b},{\bf b}'$ and write
just ${\cal U}$.
From (\ref{collins}) we have obvious

\bl\label{vanishing}
 ${\cal U}$ necessarily vanishes in one of the following cases

(A) ${\cal A}({\cal U})\neq 0$ ($d=d'$)

(B) ${\cal B}({\cal U})\neq 0$  ($d=d'$)

(C) $d\neq d'$.

\el

(iii) Next, we remind that the Schur function admit the following Taylor series
\be\label{char-map}
s_\lambda(X)=\sum_{\Delta} \phi_\lambda(\Delta)\tr\left( X \right)^{\Delta_1} \cdots
\tr\left( X \right)^{\Delta_{\ell(\Delta)}}
\ee
which is called character map relations \cite{Mac} because $s_\lambda$ is the character of
the linear group and $\phi_\lambda$ are specially normalized characters of the symmetric group
(see Appendix).
The right hand side consists of the sum of monomials of the entries of $X$.

Suppose $X=UAU^\dag B$ and the matrices $A$ and $B$ are both diagonal. Then, as one can notice
the integral over $U_N$ of each term which is monomial in the entries $X$ (this follows from the specific form of such monomials arising from traces and from the diagonality of $A$ and $B$.
For instance, $\tr UAU^\dag B = \sum U_{a,b}A_{b,b'}(U^\dag)_{b',a'}B_{a',a}$ with $a=a',\,b=b'$.)

Now suppose that $A$ is a diagonal plus a given strictly upper triangle matrix $\Delta^{(A)}$:
$\Delta^{(A)}_{i,j}=0,\,i\ge j$. For $j-i>0$ we call this number a content of the entry $\Delta^{(A)}_{i,j}$.   Then any polynomial originated from a factor
$\tr\left(UAU^\dag B\right)^m$ which contain a factor $\Delta^{(A)}_{i,j},\,i<j$ diminish
the balance number ${\cal B}$ by $j-i>0$. Each entry of $\Delta^{(A)}$ which enters a monomial ${\cal U}$ diminishes the balance number ${\cal B}$ by its content. Therefore, according to the (B) there is no contribution of $\Delta^{(A)}$
to the integral of each $\tr\left(UAU^\dag B\right)^{\mu_1}\cdots \tr\left(UAU^\dag B\right)^{\mu_k}$ and therefore, no contribution to the integral (\ref{Prop1}). Notice, that
the matrix $\Delta$ does not change the balance number ${\cal A}$.

Simirlarly, suppose the matrix $B$ is a diagonal matrix plus a strictly upper triangle matrix $\Delta^{(B)}$: $\tilde\Delta^{(B)}_{i,j}=0,\,i\ge j$. In this case, each monomial, say, ${\cal U}$  which includes any number of entries of $\Delta^{(B)}$ has the balance number ${\cal A}$ enlarges by positive number (equal to the sum of contents of the entries) and according to (A) does not contribute to the integral.

We also note that terms with $\Delta^{(A)}$ does change the balance number ${\cal A}$ and
terms with $\Delta^{(B)}$ does not change the balance number ${\cal B}$.

Since, any complex matrix can be presented in form of diagonal plus strictly triangle by the
conjugation by unitary ones (the Schur decomposition) and also due to the invariance property of the Haar measure we
obtain the proof of (\ref{Prop1}) for any complex $A$ and $B$.

As a result we have

\bl
Let $H_i=U_i X_i U_i^\dag$ where $X_i=\diag(x^{(i)}_a),\,a=1,\dots,N$ and $U_i\in\mathbb{U}_N$.
We have
\be
\int_{\Sigma^{\times n}}s_\lambda(H_1\cdots H_n)\prod_{i=1}^n d\mu(H_i)=
\prod_{i=1}^n \left(s_\lambda(I_N)\right)^{1-n}\left(\langle s_\lambda(H)\rangle \right)^n.
\ee
\el
For the proof step by step we integrate over $U_1$, then over $U_2$ and so on up to $U_{n-1}$,
taking into account that
\be
d\mu(H)=C_N\prod_{i<j}(x_i-x_j)^2 \prod_{i=1}^N e^{-\frac N2 x_i^2}dx_i d_*U,\quad
\int d\mu(H)=1
\ee
which says that the integration over unitary group can be done separately of the integration
over the eigenvalues of $H$.

Then by the Cauchy-Littlewood relation
\be\label{CLittlewood}
e^{\sum_{m=1}^\infty \frac {p_m}{m}\tr A^m}=\sum_{\lambda\atop\ell(\lambda)} s_\lambda(\pb)
s_\lambda(A)
\ee
gives rise to the following matrix models.

\subsection{Matrix models as perturbation series}

We obtain
\bp\label{Product-matr-model}
\be\label{hhh-integral}
\int_{\Sigma^{\times n}} e^{N\sum_{m=1}^\infty \frac{p_m}{m} \tr\left(H_1\cdots H_n  \right)^m}\prod_{i=1}^n d\mu(H_i)
=\sum_{\lambda\atop\ell(\lambda)\le N} s_\lambda(I_N) s_\lambda(N\pb)
\left(\frac{\langle s_\lambda(H) \rangle_{\Sigma}}{s_\lambda(I_N)}\right)^n
\ee
where $N\pb=(Np_1,Np_2,\dots)$ and
\be
\langle s_\lambda(H) \rangle_{\Sigma}=\int_\Sigma s_\lambda(H)d\mu(H)
\ee
and the exponential is considered to be formal Taylor series in the variables $p_1,p_2,\dots$.
\ep

\begin{corollary}
One can rewrite Proposition \ref{Product-matr-model}
\be\label{cor}
J_N(\pb):=\langle e^{N\sum_{m=1}^\infty \frac{p_m}{m}\tr(H_1\cdots H_n)} \rangle_{\Sigma^{\times n}}=
\sum_\lambda (N)_\lambda s_\lambda(N\pb)s_\lambda(0,1,0,0,\dots)
\left(\frac{s_\lambda(0,1,0,0,\dots)}{s_\lambda(1,0,0,0,\dots)}  \right)^{n-1}.
\ee
\end{corollary}

Here we use the known expression (see, for example \cite{Acta},\cite{HO-tmp2003})
\be
\int_\Sigma s_\lambda(H) d\mu(H)= (N)_\lambda \frac{s_\lambda(0,N,0,0,\dots)}{s_\lambda(1,0,0,0,\dots)}
\ee
obtained in \cite{HO-tmp2003} as follows. Consider the following matrix model \cite{HO-Lecce}
\be\label{2MM}
\int e^{N\sum_{m=1}^\infty
\ \left(\frac{p_m}{m}\tr H_1^m + \frac{\tilde p_m}{m}\tr H_2^m\right)
+\sqrt{-1}N\tr (H_1H_2)} dH_1 dH_2.
\ee
It is equal to 
\be\label{2MM-peturbation-series}
\sum_{\lambda\atop\ell(\lambda)\le N}(N)_\lambda s_\lambda(N\pb) s_\lambda(N\tilde\pb).
\ee
Then we take into account
\be
\int e^{- \frac{N}{2}\tr H_2^2+\sqrt{-1}N\tr (H_1H_2)}  dH_2 = e^{-\frac N2\tr H_1^2}
\ee
and obtain \cite{Acta},\cite{HO-tmp2003}
\be
\int e^{N\sum_{m=1}^\infty \frac{p_m}{m}\tr H_1^m } e^{-\frac N2\tr H_1^2}dH_1
=\sum_{\lambda\atop\ell(\lambda)\le N}(N)_\lambda s_\lambda(N\pb) s_\lambda(0,N,0,0,\dots)
\ee
and therefore we have
\be
\langle s_\lambda(H_1) \rangle_\Sigma = (N)_\lambda s_\lambda(0,N,0,0,\dots)
\ee
which finishes the proof of (\ref{cor}).

In the similar way we obtain
\bp
\be\label{cor}
J_N(\pb,C):=\langle e^{N\sum_{m=1}^\infty \frac{p_m}{m}\tr(H_1C_1\cdots H_nC_n)} \rangle_{\Sigma^{\times n}}=
\sum_\lambda s_\lambda(C_n\cdots C_1) s_\lambda(\pb)
\left(\frac{s_\lambda(0,1,0,0,\dots)}{s_\lambda(1,0,0,0,\dots)}  \right)^{n}
\ee
where the coupling constants $p_1,p_2,\dots$ and the source matrices $C_1,\dots,C_n$ are free
parameters of the model.
Here, the exponential is treated as the Taylor series in the variables $p_1,p_2,\dots$.
\ep

Let us note the we have a kind of a ``gauge freedom'': the answer depends only of the spectrum
of the product $C_n\cdots C_1=C$.

\br
Let $\deg p_m=m$.
The formal series in $\pb$ in the left hand side of (\ref{cor}) can be chosen to be a polynomial,
say $P$
if we take
$$
p_m=-\sum_i^K y_i^m
$$
where $y_i\in\mathbb{C}$ and $K$ defines the degree of the highest degree monomial terms in $P$.
\er

\subsection{Brickwork Hurwitz numbers}

Let us take $\deg p_i=i$ and $\pb_\mu=p_{\mu_1}p_{\mu_2}\cdots$, where $\mu$ is a partition.
Then $\deg \pb_\mu=|\mu|$.

Let $\pb_\kappa(C):=\tr(C^{\kappa_1})\tr(C^{\kappa_2})\cdots$, where
$C=C_n\cdots C_1$  and $\kappa$ is a partition of
$2k$ ($k=1,2,\dots$).
Using Frobenius formula (see Appendix \ref{Hu}) we obtain
\be\label{Hurwitz-bicketwork'}
J_N(\pb,C) = \sum_{k\le \frac N2}\sum_{\mu\atop |\mu|=2k} \sum_{\kappa\atop |\kappa|=2k}
{\cal H}_{S^2}\left(\kappa,\mu,\underbrace{(2^k),\dots, (2^k) }_{n} \right)
\pb_\kappa(C)\pb_\mu + O_N(\pb)
\ee
and if $C=C_n\cdots C_1=I_N$ we obtain
\be\label{Hurwitz-bicketwork}
J_N(\pb) = \sum_{k\le \frac N2}\sum_{\mu\atop |\mu|=2k} \sum_{\kappa\atop |\kappa|=2k}N^{\ell(\kappa)}
{\cal H}_{S^2}\left(\kappa,\mu,\underbrace{(2^k),\dots, (2^k) }_{n} \right)\pb_\mu +O_N(\pb)
\ee
where $\deg O_N(\pb)>N$ and
where ${\cal H}_{S^2}$ is the Hurwitz number which counts nonequivalent $2k$-fold coverings
($k\le\frac12 N$)
of the Riemann sphere with $n+2$ branch points with ramification profiles given
by partitions $\mu,\kappa,(2^k),\dots, (2^k)$. Here $\pb$ and $C$ are free parameters of the model.

Thus, the integral (\ref{hhh-integral}) is the generating function of Hurwitz numbers of certain type where the number of ramification
profiles of brick column type $(2^k)$ is equal to the number of independent Hermitian
matrices in the product which define the matrix model.

The definition of such Hurwitz numbers can be given as follows.

Consider the equation
\be\label{sphere}
X_1\cdots X_{n+2}={\rm id}
\ee
where each $X_i$ belongs to the symmetric group $S_{2k}$ ($k=1,2,\dots$) and where
elements $X_i$, $i=3,\dots,n+2$ are involutions without fixed points, and where the cycle class of
$X_1$ and of $X_2$ are given, resectively, by partitions $\kappa\vdash 2k$ and $\mu\vdash 2k$.
Then ${\cal H}_{S^2}(\kappa,\mu,(2^k),\dots, (2^k) )$ is the number of
solutions of (\ref{sphere}) divided by the number of elements in $S_{2k}$.

The model (\ref{Hurwitz-bicketwork'}) should be compared with the matrix model \cite{Eynard}
which generates simple Hurwitz numbers ${\cal H}_{S^2}(\kappa,\mu,\underbrace{(2,1^m),\dots, (2,1^m) }_{n} )$
considered by Okounkov in \cite{Okounkov}.

\section{Products of normal matrices}

Let ${\cal M}_N$ is the space of $N\times N$ normal matrices, i.e. matrices which commute
with its Hermitian conjugated. Each normal matrix can be diagonalized by  an unitary transform i.e. presented in form (\ref{Schur-decomposition}) where
$\Delta=0$.
The measure on ${\cal M}_N$ can be chosen as
\be
d\mu(M)=e^{-N\frac12\tr M^2}\prod_{i,j=1}^N d^2 M_{i,j}\delta(MM^\dag-M^\dag M)
\ee
\be
=\prod_{i<j}|z_i-z_j|^2 \prod_{i=1}^N e^{-\frac N2 |z_i|^2} d^2z_i d_*U.
\ee
We have
\bp
\be\label{normal-matrices}
\int_{{\cal M}_N} e^{N\sum_{m=1}^\infty \frac{p_m}{m}\tr \left(M_1C_1\cdots M_nC_n\right)^m}\prod_{i=1}^n
e^{N\sum_{m=1}^\infty \frac{t_m}{m}\tr \left(M_i^\dag\right)^m}d\mu(M_i)
\ee
\be\label{normal-matrices-answer}
=\sum_{\lambda\atop\ell(\lambda)\le N}s_\lambda(C)\left(s_\lambda(1,0,0,\dots) \right)^{1-n}s_\lambda(\pb)
\prod_{i=1}^n s_\lambda({\bf t}^{(i)})=
\ee
\be
\sum_{k\le \frac N2}\sum_{\mu\atop |\mu|=2k} \sum_{\kappa\atop |\kappa|=2k}
{\cal H}_{S^2}\left(\kappa,\mu,\mu^{(1)},\dots,\mu^{(n)},\underbrace{(2^k),\dots, (2^k) }_{n} \right)
\pb_\kappa(C)\pb_\mu {\bf t}^{(i)}_{\mu^{(i)}}+ O_N(\pb)
\ee
where $C=C_n\cdots C_1$.
\ep

For proof we write $M_i=U_iX_iU^\dag_i$ ($i=1,\dots,n$ and each $X_i$ is diagonal) and step by step evaluate
integrals over $U_1$, over $U_2$ and so on up to $U_{n-1}$.
Next, we take into account the Cauchy-Littlewood relation (\ref{CLittlewood}) and also
\be
C_N\int_{\mathbb{C}^{\times N}} s_\lambda(x_1,\dots,x_N)s_\mu(\bar{x}_1,\dots,\bar{x}_N)
\prod_{i<j}|x_i-x_j|^2\prod_{i=1}^N e^{-N|x_i|^2} d^2 x_i = (N)_\lambda\delta_{\lambda,\mu}
\ee
we obtain (\ref{normal-matrices-answer}). 

\br
However this result follows from \cite{NO} because normal matrices can be treated as particular
case of complex ones (however this subset does not form a group), then the basic equalities
used in \cite{NO},\cite{AOV}, which are
\be
\int_{Mat_{N\times N}(\mathbb{C})} s_\lambda(ZAZ^\dag B)d\mu(Z)=\frac{s_\lambda(A)s_\lambda(B)}{s_\lambda(1,0,0,\dots)}
\ee
and
\be
\int_{Mat_{N\times N}(\mathbb{C})} s_\lambda(ZA)s_\kappa(Z^\dag B)d\mu(Z)=\delta_{\lambda,\kappa}\frac{s_\lambda(AB)}{s_\lambda(1,0,0,\dots)}
\ee
are correct for normal matrices:
\be
\int_{{\cal M}_N} s_\lambda(MAM^\dag B)d\mu(M)=\frac{s_\lambda(A)s_\lambda(B)}{s_\lambda(1,0,0,\dots)}
\ee
and
\be
\int_{{\cal M}_N} s_\lambda(MA)s_\kappa(M^\dag B)d\mu(M)=\delta_{\lambda,\kappa}\frac{s_\lambda(AB)}{s_\lambda(1,0,0,\dots)}.
\ee
Therefore, we can construct multimatrix ensembles with random normal matrices in the same way
as it was previousely done in \cite{NO}, \cite{AOV}.
\er

\section{Tau functions}

The series (\ref{normal-matrices-answer}) can be identified with a Toda Lattice(TL) tau function

$$
\partial_{p^{(1)}_1}\partial_{p^{(2)}_1} \phi_n=e^{\phi_{n-1}-\phi_n}-e^{\phi_{n}-\phi_{n+1}},
\quad \phi_n=\frac{\tau_n}{\tau_{n+1}}.
$$

Hypergeometric tau functions solves not only bilinear but also linear equation,
which can be called string equations and which generalize Gauss equation for
Gauss hypergeometric function ${_2}F_1$.

There are a few determinantal representations of these tau functions.
For the various properties of these tau functions see \cite{KMM},\cite{OS2000},\cite{OS-TMP}.

In case any chosen pair of sets among $n+1$ sets $\pb,{\bf t}^{(1)},\dots,{\bf t}^{(n)}$, say
$\pb, {\bf t}^{(1)}$ or say ${\bf t}^{(1)},{\bf t}^{(2)}$ are treated as TL higher times (where
TL discrete variable is identified with the matrix size $N$), while the rest part of the sets
(respectively, either ${\bf t}^{(2)},\dots,{\bf t}^{(n)}$
or $\pb,\dots,{\bf t}^{(3)},\dots,{\bf t}^{(n)}$
takes
the form of

\begin{itemize}
 \item either $(1,0,0,\dots)$
 \item or $(a_i,a_i,\dots)$ any $a_i\in \mathbb{C}$, $i=1,2,\dots$
\end{itemize}

This is a tau function of hypergeometric type \cite{OS2000}.

In contrast to the integral (\ref{old}) who was recognized to be tau function in \cite{GMMMO}
the integral (\ref{new}) is related to KP tau function only in the simplest
case $n=2$.

\section{Discussion}

\begin{itemize}
 \item The models (\ref{new}) and (\ref{new'}) should be comparaed with \cite{AmbChekhov2014} and the model (\ref{normal-matrices}) should be compared with \cite{NO}.
 \item The chain matrix models can be considered in a similar way
 \item It is not clear is it possible to apply the topological recursion method to the considered models except the cases where the integrals are hypergeometrical tau functions (in this case
 one can apply the results \cite{HarnadAl}, \cite{Kazaryan}.
\end{itemize}

It will be done in a more detailed text.


\appendix

\section{Characteristc map and the Frobenius relation \label{Hu}}

The characteristic map relation expresses characters  $s_\lambda$ of the linear group
in terms of the characters $\chi_\lambda$ of the symmetric group as follows:
\be
s_\lambda(\pb)= \sum_{\mu\atop |\mu|=|\lambda|} \frac{1}{z_{\mu}}\chi_\lambda(\mu)\pb_\mu
\ee
where $z_\mu=\prod_{} i!i^{m_i}$ for $\mu=1^{m_1}2^{m_2}\cdots$. Let
$\varphi_\lambda(\mu)=\frac{\chi_\lambda(\mu)}{\chi_\lambda\left((1^{|\lambda|})\right)}{|C_\mu|}$
is the normalized character of $S_{|\lambda|}$,
where $|C_\mu|=z_\mu$ is the cardinality of the class $\mu$ and 
$\chi_\lambda\left((1^{|\lambda|}\right)$ is the dimension of the representation $\lambda$.

Frobenius formula expresses Hurwitz numbers in terms of the normalized characters:
\be
{\cal H}_\Omega(\mu^{(1)},\dots,\mu^{(n)})=
\sum_\lambda \left(\frac{\dim\lambda}{|\lambda|!}\right)^{E(\Omega)}
\varphi_\lambda(\mu^{(1)})\cdots \varphi_\lambda(\mu^{(n)})
\ee
where $E(\Omega)$ is the Euler characteristic of the closed surface $\Omega$ without boundary,

\end{document}